\documentclass[12pt,preprint]{aastex}
\usepackage{psfig,lscape}
\begin{document}

\title{The Radial Structure of SNR N103B}

\author{Karen T. Lewis\altaffilmark{1,4}, David N. Burrows\altaffilmark{1}, John P. Hughes\altaffilmark{2}, Patrick O. Slane\altaffilmark{3},
Gordon P. Garmire\altaffilmark{1}, \& John A. Nousek\altaffilmark{1}}

\altaffiltext{1}{Department of Astronomy and Astrophysics, The Pennsylvania State University, 525 Davey Laboratory, University Park, PA 16802, USA}

\altaffiltext{2}{Department of Physics and Astronomy, Rutgers University, 136 Frelinghuysen Road, Piscataway, NJ 08854, USA}

\altaffiltext{3}{Harvard-Smithsonian Center for Astrophysics, 60 Garden Street, Cambridge, MA 02138, USA}

\altaffiltext{4}{\tt lewis@astro.psu.edu}

\begin{abstract}

We report on the results from a {\it Chandra} ACIS observation of the
young, compact, supernova remnant N103B. The unprecedented spatial
resolution of {\it Chandra} reveals sub-arcsecond structure, both in
the brightness and in spectral variations. Underlying these
small-scale variations is a surprisingly simple radial structure in
the equivalent widths of the strong Si and S emission lines. We
investigate these radial variations through spatially resolved
spectroscopy using a plane-parallel, non-equilibrium ionization model
with multiple components. The majority of the emission arises from
components with a temperature of 1 keV: a fully ionized hydrogen
component; a high ionization timescale (n$_{e}$t$ > 10^{12}$
s$\;$cm$^{-3}$) component containing Si, S, Ar, Ca, and Fe; and a low
ionization timescale (n$_{e}$t$\sim$10$^{11}$ s$\;$cm$^{-3}$) O, Ne,
and Mg component. To reproduce the strong Fe K$\alpha$ line, it is
necessary to include additional Fe in a hot ($> 2$ keV), low
ionization (n$_{e}$t$\sim$10$^{10.8}$ s$\;$cm$^{-3}$) component. This
hot Fe may be in the form of hot Fe bubbles, formed in the radioactive
decay of clumps of $^{56}$Ni. We find no radial variation in the
ionization timescales or temperatures of the various
components. Rather, the Si and S equivalent widths increase at large
radii because these lines, as well as those of Ar and Ca, are formed
in a shell occupying the outer half of the remnant. A shell of hot Fe
is located interior to this, but there is a large region of overlap
between these two shells. In the inner 30\% of the remnant, there is a
core of cooler, 1 keV Fe. We find that the distribution of the ejecta
and the yields of the intermediate mass species are consistent with
model prediction for Type Ia events.
\end{abstract}

\keywords{Galaxies:Individual (Large Magellanic Cloud) --- ISM:Individual (N103B) --- ISM:Supernova Remnants --- X-Rays:ISM}

\section{INTRODUCTION}

N103B is one of the brightest radio and X-ray sources in the Large
Magellanic Cloud (LMC). As a result, this young, compact, supernova
remnant (SNR) has been selected for further study in many radio, X-ray,
and optical surveys of the LMC SNRs \citep{DM95,H95,rd90,W99} and the
general properties of this object are quite well known. At radio and
X-ray wavelengths, the emission arises from a region 30$''$ in
diameter (or 7.3 pc, assuming a distance of 50 kpc to the LMC) with
the emission in the western half being $\sim 3$ times brighter than in
the eastern half \citep{DM95,W99}. A deep H$\alpha$ image
\citep{W99,S98} reveals several bright clumps which are located in the
vicinity of the bright radio and X-ray regions. In all three bands, a
partial shell is plainly visible.

The remnant is located on the northeastern edge of an H~II region,
approximately 40~pc from the young, rich star cluster NGC~1850. While
SNRs in the Magellanic Clouds are commonly associated with H~II
regions, no other known LMC remnant is associated with a young star
cluster \citep{Chu88}. For years, it was naturally assumed that the
progenitor of N103B was a massive member of this cluster. However, the
{\it ASCA} spectrum \citep{H95} of this remnant shows strong emission
features from highly ionized Si, S, Ar, Ca, and Fe, while K-shell
emission from O, Ne, and Mg are relatively weak; this spectrum is more
consistent with the nucleosynthesis products of a Type Ia SN
\citep{N84,I99} than the core collapse of a massive star \citep{T95}.  
On the other hand, the optical spectrum is not dominated by the strong
Balmer lines often associated with the remnants of Type Ia events
\citep{Balmer82, H95}. Thus the classification of this remnant is
still somewhat uncertain.

N103B is believed to be only $\sim 1000-2000$ years old \citep{H95},
and the X-ray emission is still dominated by the ejecta. The deep,
high resolution ACIS observation provides an opportunity to study not
only the abundances but also the distribution of the ejecta, and to
compare them with models and observations of Type Ia and Type II
remnants. There are several striking differences in the ejecta
profiles of Type Ia and II SNRs, which we discuss below. By studying
the ejecta in N103B, we have gain further information about its progenitor
state. Moreover, assessing whether or not N103B is the
result of a Type Ia explosion bears on the relative number of young Ia
and non-Ia SNRs in the LMC which, though based on sparse statistics,
may be anomalous \citep{H95}

The ejecta in Type Ia SNRs may retain some of the initial
stratification generated in the explosion. As predicted by \citet{N84}
and \citet{I99}, Fe should initially remain in the interior, while Si,
S, Ar, and Ca should appear primarily near the rim of the remnant. The
{\it ASCA} observations of Tycho's SNR, which is commonly believed to
be the result of a Type Ia SN event, showed that the radial profile of
the Fe K$\alpha$ line peaks at a smaller radius than the profiles of
the other emission lines and the continuum emission
\citep{H97}. Furthermore, \citet{Hwang98} find that the Fe K$\alpha$ line is 
quite strong compared to the Fe L-shell emission. To reproduce the
correct ratio of Fe K to Fe L emission, a large amount of Fe must
exist in a very hot ($\ge 2$ keV) plasma with a low-ionization
timescale which produces primarily Fe K-shell emission. The authors
argue that this Fe is hotter than the rest of the ejecta and has a
lower ionization timescale because it is confined to the interior of
the remnant and has been more recently shocked.

By contrast, in several of the Type II SNRs in the Milky Way, for
example Cas A \citep {HCasA,Hwang00} and G292.0+1.8 \citep{P02}, the
abundance structure is quite complex and the original distribution is
not readily apparent.  Equivalent width maps of G292.0+1.8 show a
high degree of non-radial structure in which the abundances of the
ejecta in the clumps of emission vary throughout the remnant
\citep{P02}.  In Cas A, the Fe ejecta are even found {\it exterior} to
the lighter elements, indicating that the ejecta have actually
over-turned \citep{HCasA}.  

Utilizing the superb spatial resolution of {\it Chandra}, we can
observe directly whether the ejecta in N103B have retained their
original radial stratification, or whether it has been destroyed.
While the distribution of the ejecta will not likely provide a
definitive classification, it is interesting to compare this remnant
to other Type Ia and Type II remnants.

To study the distribution of the ejecta in this complex remnant, we
use a combination of narrow band imaging and spatially resolved
spectroscopy, both of which are available for the first time at the
required angular resolution with {\it Chandra}. In \S2, we describe
the data set and reduction.  The methods used to perform the narrow
band imaging and spatially resolved spectroscopy, as well as the
scientific results, are presented in \S3. In \S4, we propose a 3D
model for the distribution of the ejecta, consider their origin, and
estimate their masses.  Finally, in \S5, we summarize our findings and
suggest avenues for future work on this remnant.\

\section{OBSERVATIONS AND DATA REDUCTION}

The {\it Chandra} ACIS observation of N103B was carried out on 1999
December 4 for 40.8~ks.  At that time, the focal plane temperature of
the instrument was -110 ${^o}$C.  The remnant was positioned at the aim
point of the ACIS-S3 chip, which is one of two back-side illuminated
devices. In this paper we present the analysis of data which were
re-processed by the {\it Chandra X-ray Center} ({\it CXC}) on 2000
July 10. We processed the data using a combination of the {\tt CIAO
2.1} software package developed by the Chandra X-ray Center and tools
developed by the {\it ACIS} team at Penn State.\

The Level 1 events provided by the {\it CXC}, which include all
telemetered event grades, were first corrected for Charge Transfer
Inefficiency (CTI) as described by \citet{T01a}.  The {\it ACIS}
back-side illuminated detectors suffer from CTI in both the parallel and
serial directions, with the main effect of creating a non-uniform gain
across the device. There is also a degradation in energy resolution
which increases towards the top of the device. Both of these effects
complicate the spectral analysis of extended objects such as SNRs. For
a more detailed discussion of the CTI in {\it ACIS}, the reader is directed
to \citet{T01a}.

The data were then cleaned, following the standard {\it CXC} pipeline
threads, to remove events flagged by the {\it CXC} as well as times with a
poor aspect solution. Finally, the data were filtered to keep only
{\it ASCA} event grades (g02346). We employ response matrices created
to match the CTI-corrected data \citep{T01b}. The primary effect of
CTI-correction in this data set is to improve the spectral resolution
and create a uniform gain across the chip. The increase in count rate
is minimal so we continue to use the quantum efficiency data provided
by the {\it CXC}.

It was found that nearly 35\% of the frames exhibited intense flares
in which the background count rate over the entire detector increased
by an order of magnitude. We found no way to distinguish these
particle flares from real events, either by energy, event grade or
more sophisticated filtering methods similar to those used to remove flaring
pixels.  While a higher background count rate increases the noise in
the background-subtracted spectrum, removing the contaminating flares
would result in a significant loss of real data.  N103B is quite
compact, extending over only 30$''$, thus the contamination from this
background is quite minimal, contributing only 0.5\% of the total
counts in the source region. We find no significant difference between
the best-fit parameters and reduced chi-squared of data which include
the flares and those in which the high background frames were removed.
Therefore, in order to have as large a dataset as possible, we did not
remove the frames with background flares.

\section{RESULTS AND ANALYSIS}

\subsection{Basic Results}

\begin{figure}
\epsscale{1.0}
\plotone{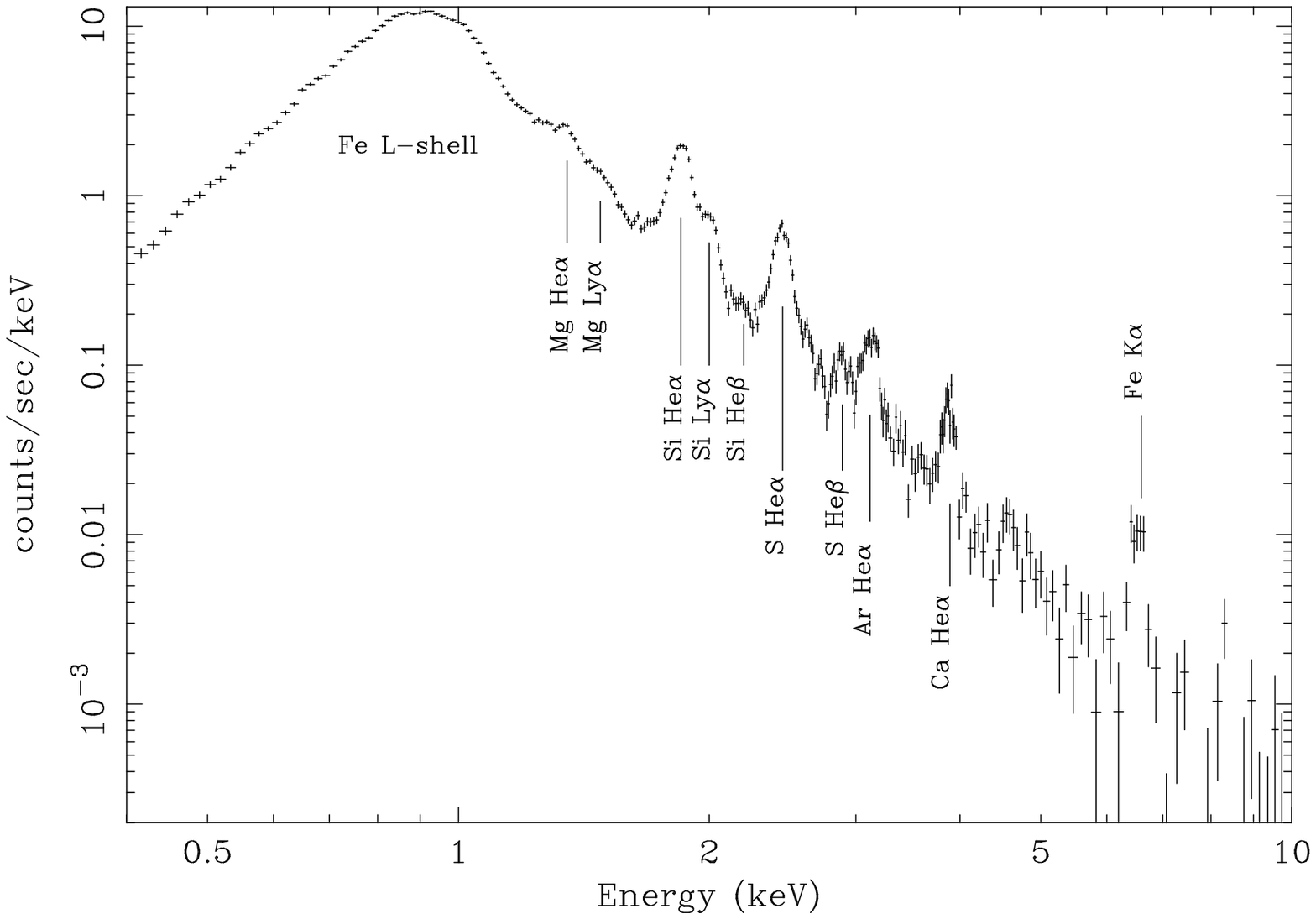}
\caption{\footnotesize \label{spectrum} The integrated spectrum for 
N103B.  The spectrum has been binned so that a minimum of 25 counts
are in each energy bin.  The K$\alpha$ and K$\beta$ lines of highly ionized
Mg, Si, S, Ar, Ca and Fe are labeled.}
\end{figure}

\begin{figure}[!t]
\epsscale{0.5}
\plotone{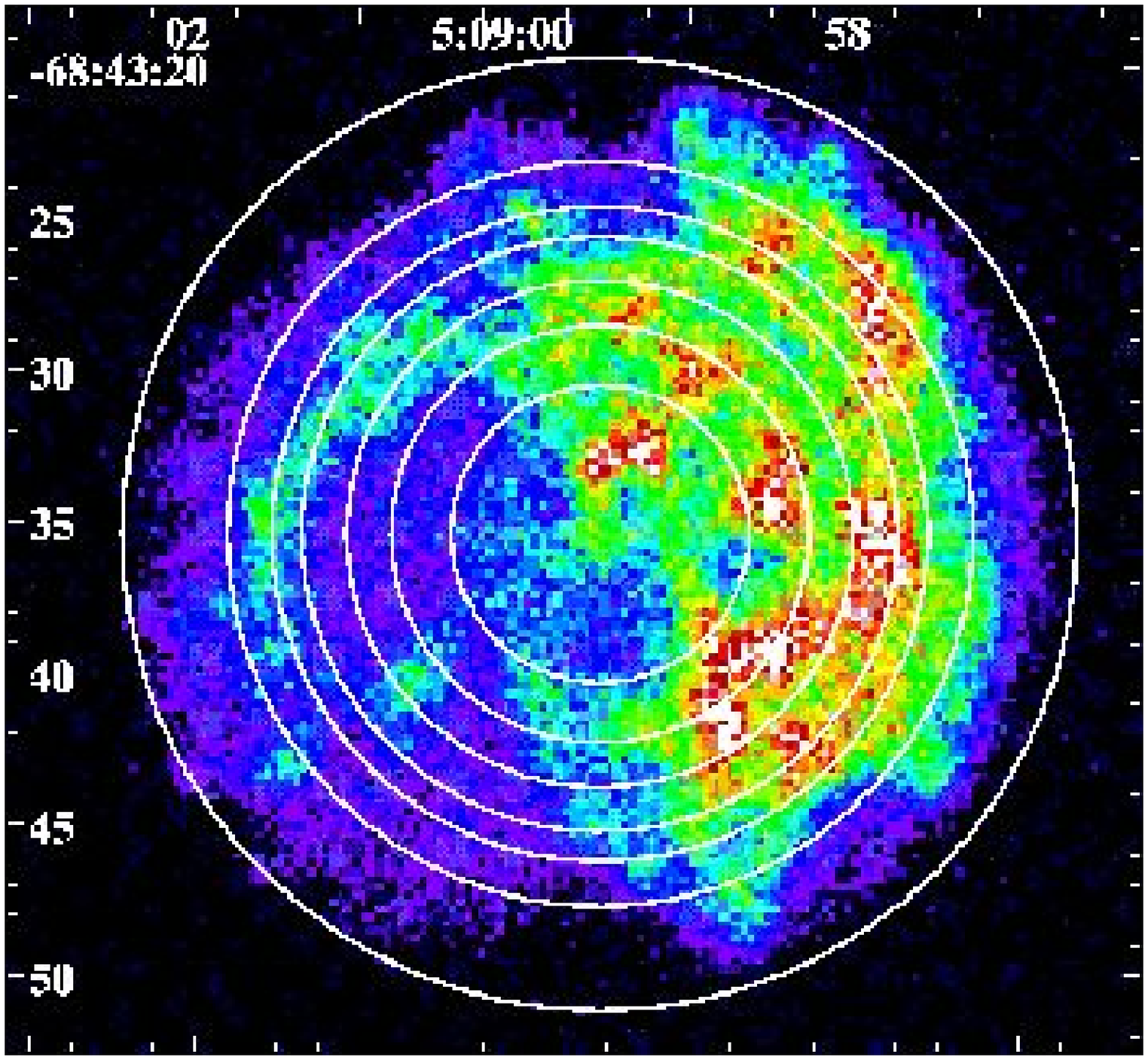}
\plotone{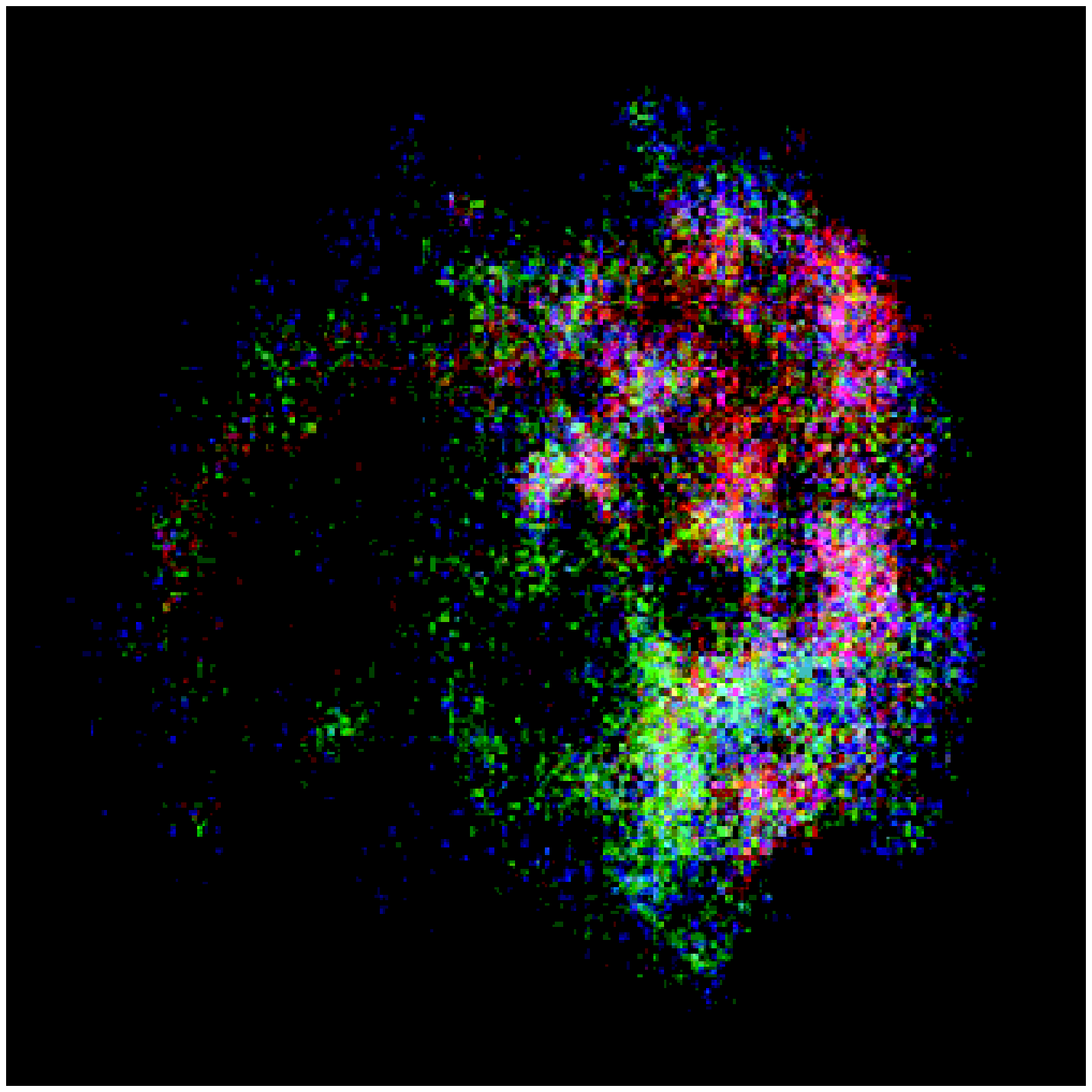}
\caption{\footnotesize \label{images} Top -- Intensity map of N103B, using
a linear scaling, ranging from 0 (black) -- 75 (white)
counts$\;$pix$^{-1}$. The annular regions to be analyzed in \S \ref{modeling}
are indicated with solid rings. Bottom -- An X-ray color image of N103B
with red (0.5 -- 0.9 keV), green (0.9 -- 1.2 keV) and blue (1.2 --
10.0 keV). The bin size in both images is 0.5$''$.}
\end{figure}

The spectrum extracted from the entire remnant is very similar to the
{\it ASCA} spectrum \citep{H95}, showing strong K$\alpha$ lines of Si,
S, Ar, Ca, and Fe and soft X-ray emission ($\sim 1$ keV) dominated by
emission from Fe L-shell transitions (Figure \ref{spectrum}). The Fe
K$\alpha$ line at $\sim$6.4 keV is much more significant in the {\it
Chandra} spectrum, as the {\it ASCA} spectrum was contaminated by
background at high energies \citep{H95}.  In Figure \ref{images}
(Top), we show the 0.4-8 keV image of N103B obtained with {\it
Chandra}.  This image reveals significant structure at the arc-second
level and the bright western half of the remnant is now resolved as a
series of dense knots as well as diffuse emission. An X-ray color
image composed from red (0.5 -- 0.9 keV), green (0.9 -- 1.2 keV) and
blue (1.2 -- 10 keV), shown in Figure \ref{images} (Bottom), further
reveals that the spectral characteristics vary throughout the remnant
as well. On large scales, the emission in the northwest is softer than
the emission from the south.  The color varies slightly from one knot
to the next, showing that the spectral variability extends to small
scales as well.

\subsection{Narrow Band Imaging}

To gain insight into the distribution of Si and S throughout the
remnant, we produced continuum-subtracted (CS) and equivalent-width
(EW) images following the methods outlined by \citet{Hwang00}. To do
this, images were extracted from the energy interval of the Si and S
emission lines as well as two continuum regions straddling each of
these lines. The energy intervals and image properties used for each
image are listed in Table \ref{ldata}.

\begin{figure}[!ht]
\epsscale{1.0}
\plotone{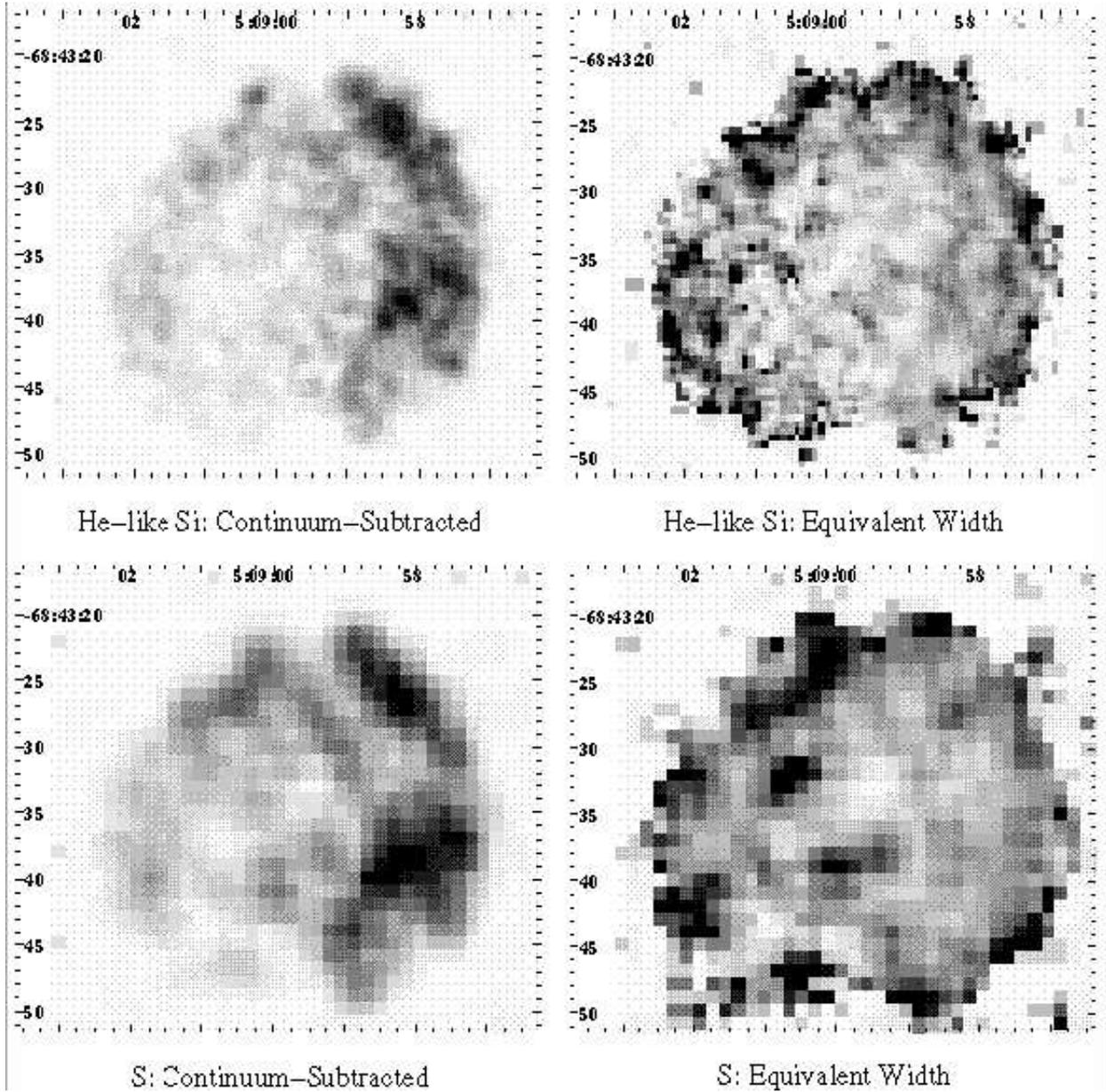}
\caption[lineimages]{\footnotesize \label{lineimage} 
Continuum-Subtracted (left) and Equivalent Width (right) images of
N103B in He-like Si and S.  All images are plotted with a linear scale
ranging from 0 (white) -- 10 counts$\;$pix$^{-1}$ (subtracted images)
and 0 (white) -- 1 keV (EW images). The bin size and smoothing size for each
image is listed in Table \ref{ldata}.}
\end{figure}

After smoothing each image with a boxcar, a linear interpolation
between the two continuum images was performed pixel by pixel to
create an image of the continuum contained in the energy interval of
the emission line. We neglected any spectral variations within the two
continuum bands, since they are negligible compared to the variations
between the bands. The CS image was produced by subtracting this
continuum image from the original line image. To create the EW image,
the CS image was divided by the continuum and then this ratio was
normalized by the width of the energy interval (in units of keV) to
obtain the EW image. As can be seen in Figure
\ref{lineimage}, the CS images for He-like Si and S are strikingly
similar to the 0.4 -- 8 keV image (Fig. \ref{images}, Top), showing
bright knots throughout the western half of the remnant. Underlying
radial trends in the remnant are clearly revealed in the EW map,
however.

It is apparent from the He-like Si and S EW images that the strength
of these lines (as compared to the continuum) increases significantly
along the rim of the remnant. To determine whether the emergence of
the ring-like structure is an artifact of the low count rate in the
rim, the original line and continuum images were collapsed into the
radial and azimuthal dimensions by binning the images into rings and
sectors. The CS intensity and EW were calculated for each bin, this
time employing a logarithmic interpolation between the continuum
regions, which is more suitable for the spectrum of N103B (see Figure
\ref{spectrum}). The radial and azimuthal plots, shown in Figure
\ref{plots}, confirm what is seen in both the CS and EW images (Figure
\ref{lineimage}). The CS intensity peaks at a radius of $\sim 10''$
with a strong azimuthal dependence. By contrast, the EW plots show
little azimuthal dependence. Most importantly, there is a clear
increase in the equivalent width with radius beginning at a radius of
$\sim 10''$ and peaking at $\sim 15''$, which is the rim of the remnant.

The Si and S EW images reveal an extremely simple radial structure,
in contrast to the spatial and spectral complexity revealed by the
false-color and X-ray color images. A detailed study of the these
radial variations present an excellent way to study the spatial
distribution of the ejecta. Thus, in the following sections, we focus
solely on the radial variations and leave an analysis of the
intriguing small-scale structure seen in the X-ray color image to
future work.

\subsection{Spectral Modeling}

To study the radial variations in more detail, we divided the remnant
into 7 concentric rings, each with $\sim 35,000 $ counts. The
extraction region for each ring is listed in Table \ref{ring_table}
and shown in Figure \ref{images} (Top).  The equivalent width (EW) of
the emission lines depends both upon the total number of emitting
atoms which are present as well as the temperature and ionization
state of the atoms. To determine the cause for the variations in the
EW, it is necessary to perform detailed spectral modeling to determine
which of these factors are contributing.

\subsubsection{The Spectral Model}\label{modeling}

We use a plane-parallel shock, non-equilibrium ionization (NEI) model
\citep{E0102}, which has been incorporated into the {\tt XSPEC}
v. 11.0.1 analysis package. This model uses a single temperature, T,
but includes a range of ionization timescales, integrated from 0 to
n$_{e}$t s$\;$cm$^{-3}$, where n$_{e}$ is the electron density and t is
the time elapsed since the material was heated to a temperature
T. This NEI model allows for each element to have independent
temperatures and ionization states which can be linked as desired.  We
use two absorbing columns, using the cross-sections of
\citet{phabs}: one to account for galactic absorption (frozen at
$7 \cdot 10^{20}$ cm$^{-2}$; Dickey \& Lockman 1990) and the other,
with lowered abundances, to model the absorption within the LMC
(allowed to vary to obtain the best fit). The abundances of the LMC
column are fixed at the values derived from the optical studies of
H~II regions and evolved supernova remnants in the LMC by
\citet{rd92} and X-ray studies of evolved supernova remnants by
\citet{H98}. When neither study provided a measurement, the abundance
of that element was set to 0.3, relative to solar, which is the
average abundance in the LMC.  Throughout, the solar abundance pattern
of \citet{anders} is used. The assumed values are given in Table
\ref{abund}. Finally, a narrow Gaussian component at $\sim 0.7$ keV is
added to all the spectra to fill in a deficit in the modeled spectrum
at the energy. A similar deficit is noted by \citet{Hwang98} in the
analysis of several SNRs using an earlier version of this NEI
code. Thus we believe that a line or line-blend is missing from the
model, and the deficit does not indicate a serious flaw in the derived
model parameters.

We find that it is impossible to fit the spectra with a single NEI
plasma model for all elemental species; it is necessary to include
several thermodynamic components to adequately model the data.  The
bulk of the continuum emission arises from a $\sim$ 1 keV plasma,
which we model as hydrogen (although any featureless continuum would
suffice as well).  Emission from Si, S, Ar, Ca, Fe also arises from a
$\sim$ 1 keV plasma with a large ionization timescale (n$_{e}$t$ >
10^{12}$ s$\;$cm$^{-3}$). In all the spectra, the strength of the Fe
K-$\alpha$ emission line is under-predicted by this single temperature
model. Following the results of \citet{Hwang98}, we allow for
additional Fe in a high temperature ($> 2$ keV) plasma which has a low
ionization timescale (n$_{e}$t$<$10$^{11}$ s$\;$cm$^{-3}$).  Finally, the
H-like Mg emission line is too weak, relative to the He-like line, to
arise from the same high ionization timescale plasma as the Si. Thus
we include O, Ne, and Mg emission in a third lower-ionization
timescale (n$_{e}$t $\sim 10^{11}$ s$\;$cm$^{-3}$) component. The
temperature of this component is poorly constrained, but is generally
consistent with 1 keV. Although Ne and O lines are not explicitly seen
in the $ACIS$ spectrum, they are seen in $XMM-Newton$ RGS \citep{vdH}
and $Chandra$ LETG \citep{M02} grating observations. In our data the O
and Ne lines are blended with the many emission lines from L-shell
transitions of Ar, Ca, and Fe, so it is not surprising that clear O
and Ne emission lines are not seen.  We add O and Ne to the
lower-ionization timescale component, as O and Ne are produced in the
O-burning zone along with Mg, as discussed in
Sec. \ref{SN_theory}. Throughout, we refer to the components as the
``H'', ``Si'',``Hot Fe'', and ``O'' components.

We do not account for any emission from Co and Ni. The bulk of the
radioactive Co and Ni produced in the initial explosion will have
decayed already, although some non-radioactive isotopes may to be
present. Furthermore, the L-shell emission of these lines overlap with
those of Fe and no K$\alpha$ lines are evident, making it difficult to
constrain the abundances of these elements. Finally, it is not
necessary to include a blastwave component at a separate temperature
to model these data. A swept-up ISM component may be related to the 1
keV hydrogen plasma that we require in our fits.

\subsubsection{Spectral Analysis Results}\label{specresults}

\begin{figure}
\epsscale{1.0}
\plotone{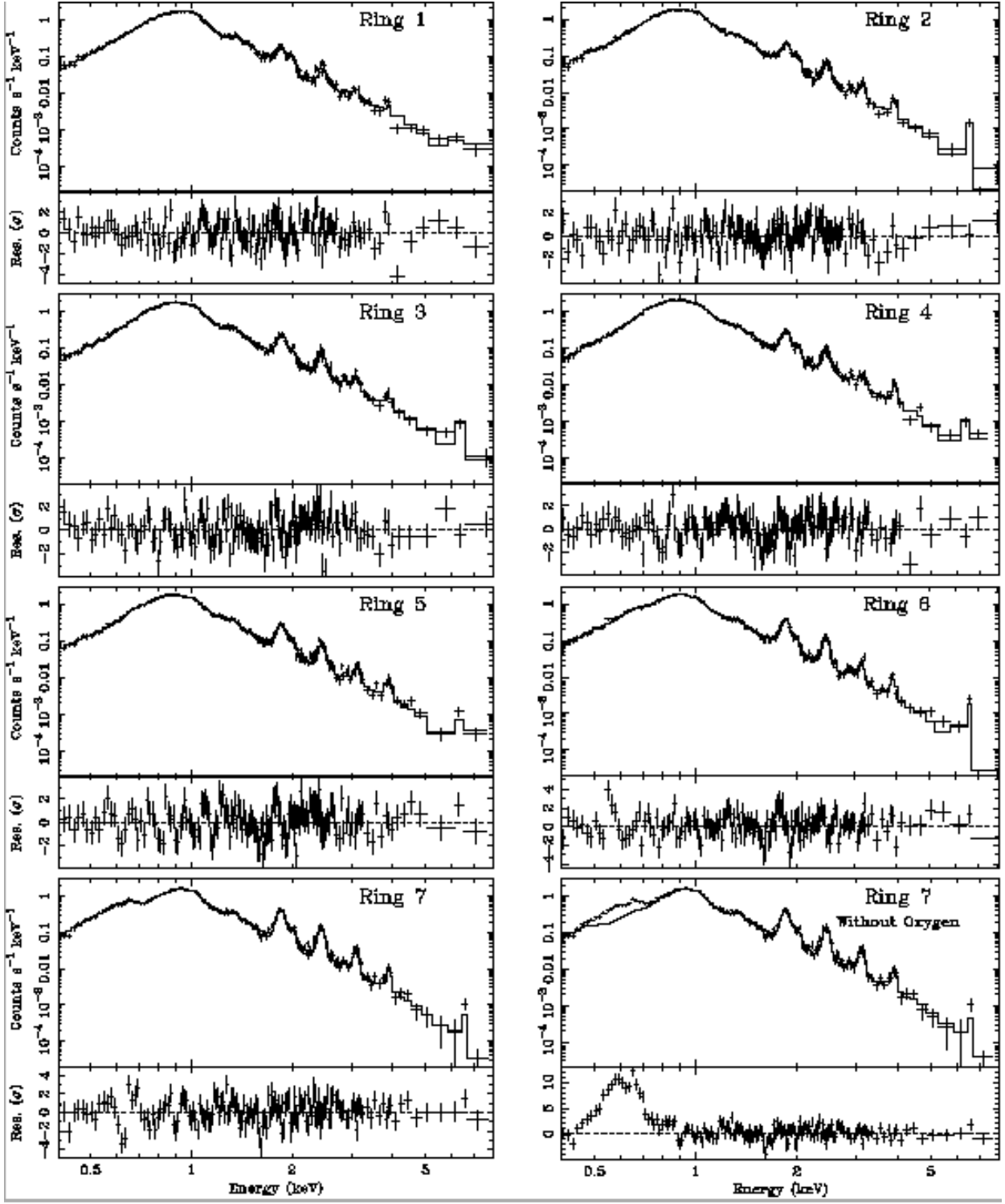}
\caption{\footnotesize \label{fits} The first seven panels 
show the spectrum from each of the seven extraction regions along with
the best fit using the four-component NEI model. The fit parameters
are shown in Tables \ref{therm_params} and \ref{abund_params}. The
eighth panel (lower right corner) shows the effect of removing the
oxygen emission from the best fit model for Ring 7. As seen in the
residuals, the oxygen emission is quite strong, despite the lack of a
clear emission line in the spectrum. The upper portion of each panel
shows the data and the model (solid line) while the bottom portion
shows the residuals, normalized to the standard deviation, $\sigma$.}
\end{figure}

The spectral fits are shown in Figure \ref{fits} and the model
parameters are listed, with 90\% errors, in Tables \ref{therm_params}
and \ref{abund_params}. These errors were obtained by stepping through
each parameter to determine the change in the $\chi^{2}$. The true
errors on the quantities may be larger, since this method assumes that
the best-fit model is correct and does not incorporate the systematic
uncertainties which may be introduced by the chosen
model. Furthermore, the error bars on quantities which are dependent
upon each other, such as temperature and ionization, may be larger
than estimated since these parameters were not varied in pairs.  Also
included in Figure \ref{fits} (lower right corner) is a plot of the
spectrum and model for Ring 7 in which the oxygen emission has been
omitted; clearly, oxygen is required to fit the data, although the
emission line is not resolved. We leave a detailed discussion of the
quality of model fits to the end of this section and here briefly
present the results of the spectral fits, which will be discussed in
more detail in the \S 4.

In Table \ref{therm_params}, we list the best fit temperatures and
ionization states for the O, Si, and Hot Fe components. The
temperature of the H component is identical to the Si temperature,
within error bars, and is not listed. In general, the temperature and
ionization timescale within each component remain fairly constant with
radius. We note that the LMC column density decreases greatly in the
outermost ring. It is difficult to determine whether this decrease
takes place throughout the entire ring, as the soft emission from the
north-west (Fig. \ref{images} Bottom) contributes a large fraction of
the total emission in the outermost ring.

In Table \ref{abund_params}, we present the volume emission measure
(EM) of each element, given by $n_{\rm e}n_{\rm i}V/4\pi D^{2}$, where
$n_{\rm e}$ and $n_{\rm i}$ are densities of the electrons and ions,
respectively, $V$ is the volume of the emitting region, and $D$ is the
distance to the remnant. Within the O and Si components we report the
EMs of the other elements in the component relative to either the O or
Si EM. If the emission from each species within a single component
arises from the same volume, then the ratio of two EMs within a
component is simply a ratio of the ion densities of the two elements,
which in turn is related to their abundances and masses. As described
in the previous section, the O, Ne, and Mg emission lines are not
clearly resolved and the EMs associated with this component should be
treated with caution.

\begin{figure}[!ht]
\epsscale{0.8}
\plotone{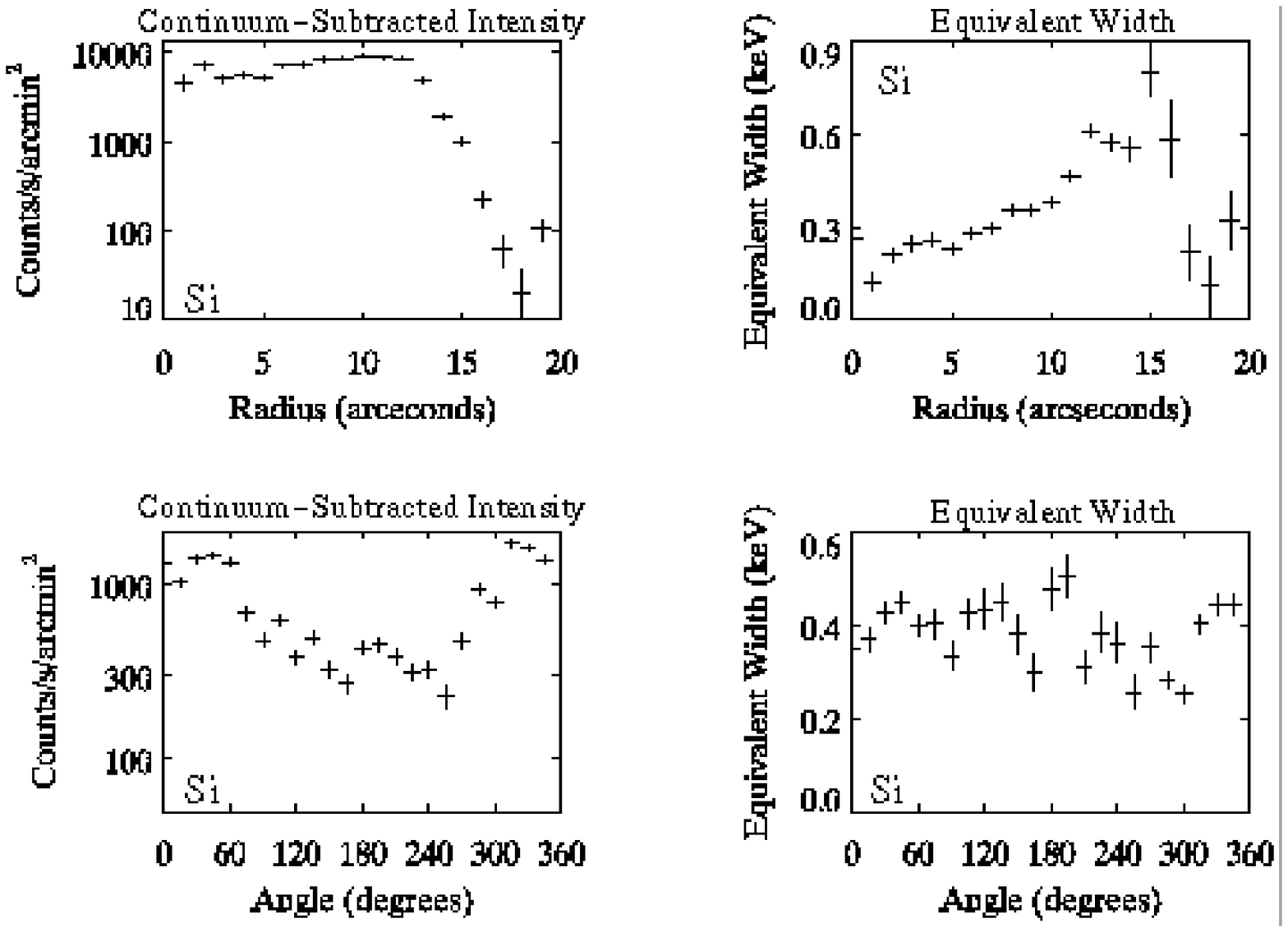}
\plotone{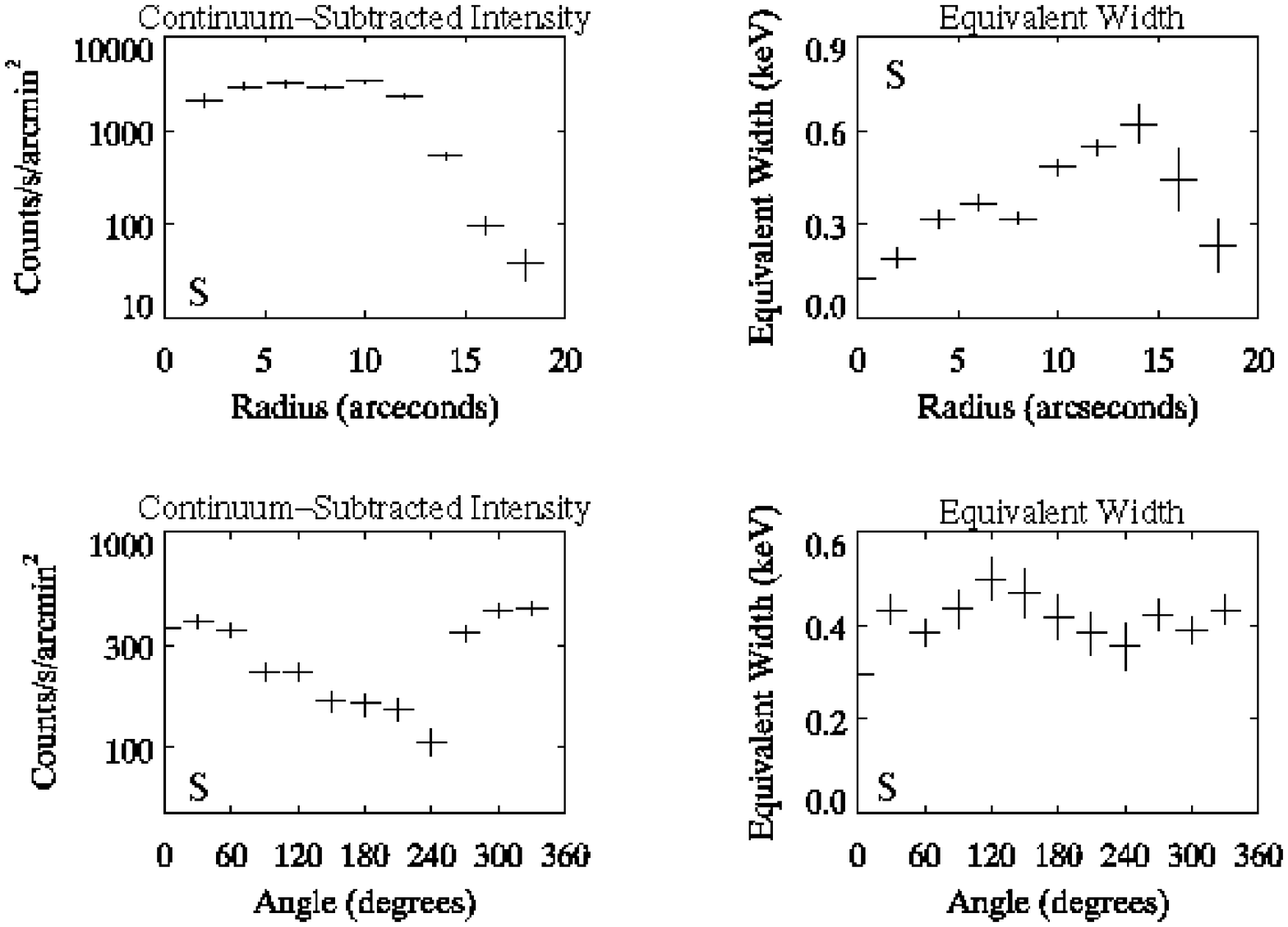}
\caption[plots]{\footnotesize \label{plots} The dependence of the 
Continuum-Subtracted Intensity (left) and Equivalent Width (right)
upon radius and angle for He-like Si (top two panels) and S (bottom
two panels). 0$^{\rm o}$ is West.}
\end{figure}

There are several striking trends in the EMs reported in Table
\ref{abund_params}. First, the EM of Si rises steadily from Ring 1 to
Ring 7, mimicking the behavior of the equivalent width of He-like Si
shown in Fig. \ref{plots}, while the EM of the hot Fe declines (See
Fig \ref{Si_Fe}).  Second, the EMs of S, Ar, Ca, and Fe, relative to
Si, are roughly constant within the error bars. An exception is the
highly elevated Fe EM in Ring 1. Third, neither the H nor O EM show
any clear radial trend and they vary in concert, with the O EM being
2$\cdot 10^{-5}$ times the H EM, within errors.  Although the Ne and
Mg EMs, relative to O, are not constant within error bars, they also
do not show any clear radial trend.

From this spectral analysis, it is clear that the increase in the
equivalent widths of Si and S with radius seen in the narrow band
images (Figs. \ref{lineimage} and \ref{plots}) is not due to a change
in the temperature or ionization state of the Si and S emitting
plasma, but an increase in the EM of the Si component.  Since the
temperature, ionization, and composition of the Si component are
roughly the same in the seven extraction regions, the only way to
increase the EM is by increasing either the ion density or the volume
of the Si component contained in the outer extraction regions. In
either case, the variations in the EM trace the {\it physical
distribution} of the Si component, and can be a powerful tool for
studying the morphology of this emission component. Similarly, the
distribution of the hot Fe component is also traced by its EM.  In
\S4.2, we model the variations in the EMs of the hot Fe and Si
components in detail to place constraints on the three-dimensional
distribution of the material in these two components.

Finally, we note that the densities of S, Ar, and Ca, relative to Si,
are elevated above those expected from the LMC (Table \ref{abund}),
while the relative density of the Fe in the Si component is
considerably lower than expected from the LMC.  Although some ISM may
have been swept up, it is clear that the emission in the Si component
arises primarily from ejecta. The hot Fe is naturally assumed to also
be a component of the ejecta. However, the densities of Ne and Mg,
relative to O, are significantly higher than expected from the LMC
(0.18 and 0.054, Table \ref{abund}) or either a Type Ia (0.014 and
0.059) or Type II (0.11 and 0.069) SNR \citep{I99,T95}. It is not
entirely clear whether the O, Ne, and Mg emission arise from ejecta or
the swept-up material. In \S 4.3 and \S 4.4, we will study the EMs in
detail to investigate the origin of the various emission components.

\subsubsection{Discussion of the Spectral Fits}

The model fit in all seven rings is generally quite good across the
entire 0.4-8 keV range. Here we discuss several sources of the
residuals and their impact upon the fit parameters.

Although the CTI correction compensates for gain variations across the
detector, minor errors ($ \le 1\%$) in the gain are expected to remain
\citep{T01b}.  This is particularly true for data obtained at a focal plane 
temperature of -110$^{\rm o}$ as there is a limited data-base of
calibration data with which to tune the CTI corrector.  We noticed
slight systematic shifts in the energy centroids of the strong K-shell
Si, S, Ar and Ca emission lines, which become progressively worse in
the outer rings.  While these shift are not large, the data are of
sufficient quality that they affect the $\chi^{2}$ of the fit
dramatically Further, the emission measures of the elements would be
somewhat underfit, compromising our analysis of the ejecta
distribution. Thus, in our fits we applied a slight adjustment to the
gain ($ < 1\%$), by including the slope of the gain as an additional fit
parameter. Some slight offsets are still visible in the residuals, but
these are not significant.

The residuals in the fit to the spectrum of Ring 7 are considerably
larger than seen in the other rings. The spectra of the eastern and
western halves of this ring are quite different below 1 keV, which
makes it difficult to fit the combined spectrum.  The spectrum of the
eastern half of Ring 7 is quite similar to those of the inner six
rings while the spectrum of the western half exhibits the 0.7 keV
deficit seen in the combined spectrum (Fig. \ref{fits}).  The quality
of the spectral fit to the eastern half of Ring 7 is quite good
($\chi^{2}_{\nu} = 1.3$) with fit parameters, particularly the
emission measures, similar to those derived from the fit to Ring 7 as
a whole, although with larger error bars due to the greatly reduced
count rate.  Thus we are confident that, despite the statistically
poorer fit to Ring 7 as a whole, the derived fit parameters are
reliable.

\section{DISCUSSION - ORIGIN AND DISTRIBUTION OF THE EJECTA}

In the previous section, we demonstrated that N103B contains a large
amount of hot Fe with a low ionization timescale, similar to that
found in Tycho's SNR \citep{Hwang98}. As shown by the radial
variations in the emission measure (EM), this hot Fe component is
located centrally, as seen in Tycho's SNR \citep{H97}, while the Si
component, composed of Si, S, Ar, Ca, and additional Fe, is more
prominent in exterior regions of the remnant. Qualitatively, the
distribution of the elements is more similar to that inferred for
Tycho's SNR than that seen in Cas A and G292.0+1.8. With this in mind,
we compare the properties of N103B with remnants of Type Ia SNe.

To aid the following discussion, we begin with a brief overview of
nucleosynthesis models in Type Ia SNe and the predicted distribution
of the ejecta. Then in \S\ref{3D}, we model the variations in the Si
and hot Fe EMs in detail, taking into account the 3D structure, to
place constraints upon the distribution of the ejecta. In
\S\ref{hotFe}, we discuss the origin of the hot Fe component and in
\S\ref{HandO} briefly discuss the H and O components. Finally, in
\S\ref{mass}, we calculate the masses of the ejecta and compare with
the expected masses of the ejecta produced in Type Ia and Type II SNe.

\subsection{Overview of Nucleosynthesis in Type Ia SNe}\label{SN_theory}

\citet{I99} model the nucleosynthesis in Chandrasehkar mass models 
for Type Ia SNe.  The authors consider pure deflagration models (of
varying deflagration speed) as well as delayed detonation models in
which the deflagration front transitions to a detonation front.  Here
we describe the fast deflagration W7 model \citep{N84,I99}.  Complete
Si burning occurs in the inner $\sim 0.8~\rm M_{\odot}$ of the white
dwarf, producing the Fe-peak nuclei: Ni, Fe, and Co. The central core
($\sim 0.2~ \rm M_{\odot}$) is primarily non-radioactive $^{56}$Fe and
$^{54}$Fe, but $^{56}$Ni is the dominant species produced throughout
the rest of the Si-burning zone. Incomplete Si burning (from $0.8-1.2
~\rm M_{\odot}$) leaves behind $^{54}$Fe, Ca, Ar, S, and Si.  Finally,
in the outer $\sim 0.2~ \rm M_{\odot}$ O, Ne, and C are burned,
leaving behind additional Si as well as Mg, Ne, and O. The transition
from one burning zone to the next is rather sharp and the overlap
between complete and incomplete Si burning products is quite small.
The delayed detonation models produce a qualitatively similar
distribution, but the transition between the complete and incomplete
Si-burning zones is blurred and the Fe-peak nuclei and Si-burning
products can co-exist throughout a large fraction of the remnant.

\subsection{3D Distribution of the Ejecta}\label{3D}

In \S 3.3.2, we concluded that the EMs in the Si and hot Fe components
change either because the ion density or volume changed from one
extraction region to the next. In principle, one could assume that all
seven extraction regions contained an identical volume of ejecta from
both components and that the ion densities varied from one region to
the next. However, it is more physically reasonable to explain the EM
variations with changes in the volume.

As suggested by the models for Type Ia SN nucleosynthesis presented
above, in the absence of mixing, the ejecta of a SNR are expected to
be radially stratified with Fe-rich core surrounded by a Si-rich
shell. Although each of the seven extraction regions we have chosen
will include emission from both the Fe core and Si shell, an interior
extraction region will contain a larger volume of the Fe core than an
exterior extraction region, while the opposite holds for the Si
shell. Qualitatively, one can easily account for the variations in
the Si and hot Fe emission measures by this simple geometrical effect.

\begin{figure}[!ht] 
\epsscale{1.0}
\plotone{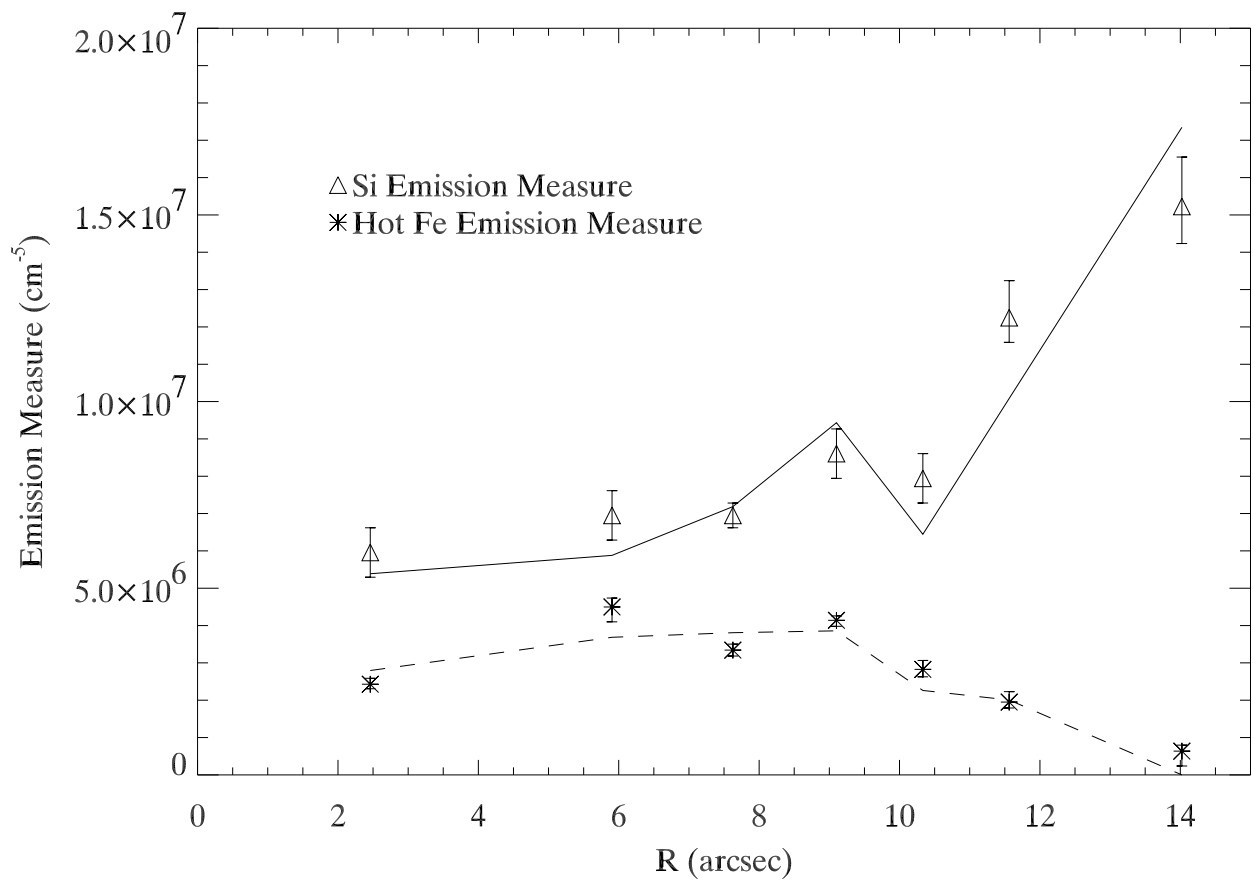}
\caption{\footnotesize \label{Si_Fe} Plot of the Si and hot Fe emission 
measures (EMs) as a function of radius. The data points and error bars
represent the fit values, shown in Table \ref{abund_params}, while the
lines connect the predicted EM for each ring, calculated using the
simple 3D model for the ejecta discussed in \S\ref{3D}. The sudden
increase in both the measured and predicted EMs in Ring 4 is due to
the slightly larger volume contained in Ring 4, as compared to Rings 3
and 5.}
\end{figure}

To place constraints upon the 3D distribution of the ejecta, we begin
by modeling the ejecta as a spherical volume with three zones. An
outer shell is filled by the Si component and hot Fe is located in a
shell interior to this. To account for the enhancement of Fe in Ring
1, we include a small sphere of Fe in the center of the remnant. This
is also consistent with the slight decrease in the hot Fe EM in Ring
1, as the volume of the hot Fe zone contained in the Ring 1 is reduced
(Table \ref{abund_params} and Fig. \ref{Si_Fe}). The inner sphere of
Fe may be associated with the core of $^{54}$Fe and $^{56}$Fe. Within
each zone, the density is assumed to be constant. The transition radii
between the zones are given by $\rho_{1}$ and $\rho_{2}$, where $\rho$
is normalized by 16$''$, which is the projected radius of the
outermost extraction region.

From this 3D model, we predict the EMs that would be observed by
determining the volume of each of the three zones which is contained
in a particular extraction region, given the assumed $\rho_{1}$ and
$\rho_{2}$. The volume of a sphere contained in a 2D annular
extraction region is given by: 

\[ V = 4\pi \int_{{\rm R}_{1}}^{{\rm R}_{2}}\rm{RdR} 
\int_{\sqrt{\rho^{2}-{\rm R}_{1}^{2}}}^{\sqrt{\rho^{2}-{\rm R}_{2}^{2}}}dz =
\frac{4}{3}\pi((\rho-{\rm R}_{1}^{2})^{3/2} - (\rho-{\rm R}_{2}^{2})^{3/2}),\] 

\noindent where R and z are the cylindrical coordinates, $\rho$ is the 
radius of the sphere, and R$_{1}$ and R$_{2}$ are the inner and outer
radii of the annular extraction region. Given the inner and outer
radius of each of the three zones, $\rho_{1}$ and $\rho_{2}$, it is
possible to determine the volume of each zone which is contained in
a particular extraction region. 

\begin{figure}[!ht] 
\epsscale{0.7}
\plotone{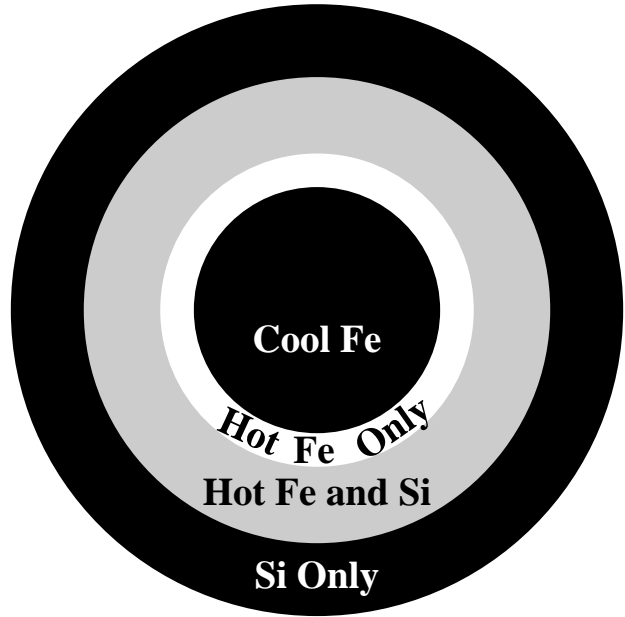}
\caption{\footnotesize \label{cross_section} Model cross-section of N103B,
based upon the three-zone model developed in \S\ref{3D} and the fit
parameters found in Table 6.}
\end{figure}

In each extraction region, EM$_{\rm obs}$ = $K\cdot V_{\rm
zone}/V_{\rm tot}$, where EM$_{\rm obs}$ is the measured EM (Table
\ref{abund_params}), $V_{\rm zone}$ is the calculated volume of the
zone which is contained in the extraction region, and $V_{\rm tot}$ is
the total volume of the remnant. In terms of physical quantities, $K$
= $n_{\rm e}n_{\rm i}V_{\rm tot}/4\pi D^{2}$ cm$^{-5}$. The Si and hot
Fe EMs were fit separately by stepping through a range of $\rho_{1}$,
$\rho_{2}$, and $K$ to map out a 3D $\chi ^{2}$ space to locate the
minimum. When fitting the Si EM, $\rho_{1}$ is set to 0, since changes
in this parameter do not affect the Si EM. The fit parameters are
shown in Table 6 and the best fits are plotted in
Fig. \ref{Si_Fe}. Although the fits are statistically unacceptable,
the qualitative agreement between the observed data and the
predictions of this simple model is remarkable.  In
Fig. \ref{cross_section}, we show a schematic diagram of the
three-zone model for N103B.

Although this model assumes that the hot Fe and Si zones do not mix,
it is clear from the different fit values of $\rho_{2}$ (Table 6) that
there must be a large region of overlap between the two zones. As a
result, the masses obtained in \S\ref{mass} will be somewhat
overestimated, because we assume throughout that each component
completely fills the volume of its zone. The large overlap between the
hot Fe and Si components could indicate that the distribution may be
more similar to that predicted by the delayed detonation models (See
\S\ref{SN_theory}). However, significant mixing between these two layers 
could occur at a later time, and we cannot rule out a fast
deflagration model for this remnant.

\subsection{Origin of the Hot Fe}\label{hotFe}

We have found that the Fe K-shell and a large fraction of the Fe
L-shell emission arises from a hot ($> 2$ keV), low ionization
timescale (n$_{e}$t$\sim$10$^{10.8}$ s$\;$cm$^{-3}$) plasma. Although the
hot Fe is located in the interior, it overlaps spatially with the Si
component. It would be very difficult to shock the hot Fe at a
significantly later time than the Si component, as suggested for
Tycho's SNR \citep{Hwang98}. Alternatively, the electron density in 
the hot Fe component could be much lower than in the Si component,
leading to a low ionization timescale.

In their analysis of SN 1987A, \citet{Li93} find that
$\sim 10$\% of the energy released in the radioactive decay sequence
$\rm{^{56}Ni(e^{+},\gamma)^{56}Co(e^{+},\gamma)^{56}Fe}$ is absorbed
locally. As a result, clumps of Ni/Co/Fe are heated to a much higher
temperature than the rest of the plasma. These clumps expand until
they come into pressure equilibrium with the surrounding plasma. It is
estimated these bubbles have a filling factor of 30$\%$ in SN 1987A.
While the hot, low density bubbles initially include Ni, Fe, and Co,
the bubbles will eventually be composed primarily of $^{56}$Fe. The
low density of the bubbles leads naturally to a lower ionization
timescale.

\citet{B01} model the dynamics of these Fe bubbles. As the
reverse shock passes through the bubbles, the postshock temperature
is significantly higher, due to the low density of the
bubbles. Additionally, they find that the presence of Fe bubbles leads
to vigorous mixing, which will destroy the initial radial structure of
the remnant; a distinct shell of ejecta are not expected.  Large
variations in pressure and temperature are expected as well.  They
further find that as the bubbles expand into the surrounding material,
the ejecta are compressed into narrow, dense filaments which partially
trace out the boundaries of the Fe bubbles. It is expected that 90\%
of the emission measure in the SNR may arise from less than 2\% of the
volume of the shocked gas.

\citet{WC01} consider the effects of Fe bubbles in the specific case
of a Type Ia remnant with an exponential density profile which is
expanding into a constant density medium. This study was motivated, in
part, by observations of Tycho's SNR, which indicate the existence of
ejecta at very large radii \citep{H97} as well as two nearly
undecelerated knots of ejecta \citep{Hu97}.  They suggest that the
ejecta at large radii are primarily in the form of dense clumps which
decelerated less rapidly than the surrounding ejecta. To exist so
close to the boundary of the shock, the knots must be $\sim$ 100
denser than the surrounding medium. This high compression factor is
not a natural consequence of Rayleigh-Taylor instabilities and
\citet{WC01} argue that Fe bubbles are {\it necessary} to create such dense
clumps of ejecta. Since the hot Fe bubbles and the other ejecta are
expected to be shocked at the same time, the ionization timescale for
the hot Fe is expected to be $\sim$ 100 smaller than for the other
ejecta.

The Fe bubbles are an attractive explanation for the hot Fe seen in this
remnant for several reasons. The location of the hot Fe component, as
shown in Fig. \ref{Si_Fe} is consistent with the expected distribution
of $^{56}$Ni, as described in \S\ref{SN_theory}.  Furthermore, the Fe
component has an ionization timescale which is $\sim$ 100 times lower
than that of the Si component.  Morphologically, N103B has a
filamentary structure, particularly on the western side of the remnant
(Fig. \ref{images}, Top) and as in Tycho's SNR, there is a large amount
of ejecta at relatively large radii (Fig. \ref{plots}). However, the
mixing in N103B does not appear to be as severe as predicted by the
hydrodynamical simulations of \citet{B01} and \citet{WC01}.

\subsection{The Hydrogen and Oxygen Components}\label{HandO}

Before beginning a discussion of the H and O components, it is
important to note that the EMs of O, Ne, and Mg should be treated with
caution, since there are no strong, isolated, emission lines from
these species observed in our spectrum (See \S\ref{modeling}). Thus,
any conclusions drawn here should be viewed with some caution.

While the EMs of the H and O components vary quite a bit from one ring
to the next, there is no clear radial trend, as is seen in the Si and
hot Fe components. The H and O components vary in concert, with the O
EM $\sim 2\cdot 10^{-5}$ times the H EM and the EMs of both components
$\sim$ 50\% larger in the rings in which the bright clumps occupy a
large projected area of the extraction region. This strongly suggests
that the H and O components must be physically connected. The lack of
radial variations and the sharp drop in emissivity beyond 16$''$
suggests that the H and O components are not in a shell surrounding
the remnant, but are found in a clumpy foreground or background
structure in the ISM which has interacted with the remnant and been
heated to X-ray emitting temperatures.

However, the number density of O, relative to H, is an order of
magnitude smaller than expected from material in the LMC. This problem
would be resolved if the O component were linked with only a portion
of the H component, but then it would be difficult for the two
components to fluctuate simultaneously by 50\%. On the other hand, the
Ne and Mg number densities, relative to O, are much higher than
expected from the ISM. 

One could argue that the O component is composed of ejecta. However,
in this case, it is very difficult to understand the intimate
relationship between the H and O components and the lack of radial
variations in the O EM. Furthermore, the number densities of Ne and
Mg, relative to O, are also higher than expected for both Type Ia and
Type II SNRs. 

Using the radial variations alone, very little can be said about the
origin or distribution of the H and O components. Spectroscopy of the
bright clumps may give more insight into these questions. It is clear,
though, that the H and O components do not occupy the same physical
space as the Si component. Therefore, the abundances and EMs of the
species in the O and Si component should be compared with extreme
caution.

\subsection{Masses of the Ejecta}\label{mass}

Using the emission measures (EMs) listed in Table
\ref{abund_params} and the 3D model of the ejecta distribution found
in \S\ref{3D}, the mass of each element can be estimated. Recall that
the normalization factor used in the 3D model is $K = n_{\rm
e}n_{\rm i}V_{\rm tot}$/4$\pi D^{2}$ cm$^{-5}$, where $V_{\rm
tot}$ is the total volume of the remnant. 

Within the Si component, the electron density, $n_{\rm e}/n_{\rm Si}$,
is given by the sum over all species of $n_{i}/n_{\rm Si}X_{i}\cdot
Z_{i}$, with $i$ ranging from Si to Fe. We obtain $n_{i}/n_{\rm Si}$
directly from Table \ref{abund_params}, $X_{i}$ is the ionization
fraction calculated within the NEI code, and $Z_{i}$ is the charge of
each species. We averaged the values of $n_{i}/n_{\rm Si}$ for S, Ar,
and Ca from all seven rings, but excluded the first ring when
calculating $n_{\rm Fe}/n_{\rm Si}$. Using this information, we
obtained $n_{\rm e}/n_{\rm Si}$ = 39. Since the extraction region of
Ring 7 extends to 16$''$, or 3.9 pc, we use this as the boundary of the
total emitting volume. Using $D$ = 50 kpc, we obtain $n_{\rm Si} =
1.4\cdot 10^{-3} {\rm cm}^{-3}$. If we assume that the Si component
completely fills the Si zone ($\rho$ = 0.51--1.0), then we obtain a
Si mass of 0.21M$_{\rm \odot}$. Using the average $n_{i}/n_{\rm Si}$
used above, we obtain the masses of all the elements in the Si
component, which are listed in Table \ref{mass_table}. 

Similarly, in the hot Fe component, $n_{\rm e}/n_{\rm hot\, \rm Fe}$ =
19 and $n_{\rm hot \, \rm Fe} = 1.6\cdot 10^{-3} {\rm
cm}^{-3}$. Assuming that the hot Fe component fills the region $\rho$
= 0.4 -- 0.77, we obtain 0.21M$_{\rm \odot}$ of hot Fe.  Finally, to
estimate the mass in the Fe core, we use the excess Fe EM in Ring 1,
which is 3.9$\cdot 10^{6}$ cm$^{-5}$. We also note since there is an
Fe enhancement only in Ring 1, the Fe core probably does not exceed
$\rho = 0.32$. In this zone, $n_{\rm e}/n_{\rm Fe}$ = 21, leading to
$n_{\rm Fe} = 5.1\cdot 10^{-3} {\rm cm}^{-3}$ and a Fe core mass of
0.06 M$_{\rm \odot}$. From all three zones, we have a combined Fe mass
of 0.34 M$_{\rm \odot}$. It is impossible to calculate the masses
in the H and O components because we have no estimate of the emitting
volume.

The masses calculated above could be over- or underestimated for
several reasons.  First, we note that the estimated electron density
in the Si component is not even twice that in the Fe component. If we
wish to rely only upon a difference in electron density to account for
the difference in ionization timescale, the Si component will need to
be in the form of dense clumps with a very small filling factor. This
is not inconsistent with the simulations performed by \citep{B01}
\citep{WC01}. However, if the Si component is highly clumped, the
masses from this component will be significantly reduced. Second, the
assumption of a pure metal plasma results in a larger mass than if we
had assumed a highly enriched but still H/He dominated plasma. This is
particularly relevant in the Si component, as it is possible that some
ISM has been swept up into this outer shell.  Finally, we cannot
assume that all of the ejecta are visible in X-rays. Since the remnant
is not center-filled, it is possible that the ejecta in the interior
have not been completely shocked by the reverse front. Thus the mass
of the Fe ejecta in particular may be considerably {\it higher} than
estimated above.

Given these considerations, the estimated masses of Si, S, Ar, Ca, and
Fe are in better agreement with the predictions of a Type Ia (W7)
model than a core-collapse SN (Table \ref{mass_table}). Additionally,
the O-rich component of N103B appears to be associated with a
structure in the ISM, rather than an O-rich zone of ejecta one would
expect from a core-collapse SN.

\section{CONCLUSIONS AND SUGGESTIONS FOR FUTURE WORK}

The spectrum of N103B obtained through this {\it Chandra} observation
is quite similar to the ASCA spectrum \citep{H95}, showing strong
K$\alpha$ lines of Si,S, Ar, Ca and Fe, suggesting a Type Ia origin
for the remnant. The {\it Chandra} image of N103B reveals
structure at the sub-arcsecond level. The bright western side of the
remnant, seen in previous X-ray images, is composed of a series of
bright knots and filaments. An X-ray color image reveals that the
spectral characteristics also vary dramatically throughout the
remnant.

We find that despite the complex spatial and spectral morphology
suggested by the false-color and X-ray color images, there are striking
radial trends in the equivalent widths of the strong Si and S emission
lines, as revealed by narrow band imaging.  The equivalent widths
remain fairly constant within the interior, then rise rapidly at a
radius of 10$''$. To investigate the cause for the increase in
equivalent width through spatially resolved spectroscopy, we divided
the remnant into seven concentric rings, each with approximately
35,000 counts.

The data are well fit by a plane-parallel, non-equilibrium ionization
model. The continuum emission arises primarily from a 1 keV hydrogen
plasma.  An additional $\sim$ 1 keV plasma with a high ionization
timescale (n$_{e}$t$ > 10^{12}$ s$\;$cm$^{-3}$) contains Si, S, Ar, Ca,
and Fe. A hot ($> 2$ keV), low ionization (n$_{e}$t$\sim$10$^{10.8}$
s$\;$cm$^{-3}$) Fe plasma is required to produce the strong Fe K$\alpha$
line. Finally, the O, Ne, and Mg are located in a plasma with an
ionization timescale of n$_{e}$t$\sim$10$^{11}$ s$\;$cm$^{-3}$ and
temperature of roughly $\sim$ 1 keV. The components are referred
to as the H, Si, hot Fe, and O components, respectively.

Using this spectral model, we have determined that there are no
significant radial variations in the temperatures or ionization
timescales of the components.  Instead, we find that the emission
measures (EM = $n_{e}n_{i}V/4\pi D^{2}$) of the species in the Si
component increase radially, mimicking the radial profiles of the Si
and S equivalent width images. An exception is an enhancement of Fe in
the innermost extraction region. In contrast to the Si component, the
hot Fe EM has a profile which drops rapidly at radii greater than
10$''$.

The EM variations in the Si and hot Fe components are well modeled by
a simple three-zone model for the ejecta. In the interior of the
remnant is a sphere of Fe with a temperature of 1 keV and a high
ionization timescale, which occupies only the inner 3\% of the
remnant's volume. Exterior to this is a shell of hot Fe which is
plausibly in the form of hot Fe bubbles.  Finally, surrounding the hot
Fe is a shell of Si, S, Ar, Ca, and Fe. The Si and hot Fe components
coexist for a large fraction of the remnant volume, implying that the
difference in ionization timescale, $n_{e}$t, is due to a difference
in electron density.

We have limited information about the location and origin of the H and
O components. The EMs of these components show no radial trends, like
those seen in the Si component.  Furthermore, these components vary in
concert, suggesting that the two components are physically linked.  It
is likely that these two components are associated with a clumpy
foreground or background structure in the ISM which has been shocked
by the remnant, rather than a shell of ejecta or swept-up material. It
is clear that the O component does not occupy the same volume as the
Si component, and that any comparison between these two components
must be performed with care.  In particular global O, Si, and Fe
abundances derived from integrated spectra of this remnant cannot be
directly compared to nucleosynthesis models without first taking into
account the different physical locations of the different components.

Finally, we estimate the masses of Si, S, Ar, Ca, and Fe and find that
they are more consistent with the yields of a Type Ia SN than a Type
II SN. In particular, the large mass of Fe (0.34 M$_{\rm \odot}$)
suggests that a Type Ia origin for N103B is more likely. Further
support for a Type Ia origin is the lack of an O-rich component of
ejecta. Finally, the properties of N103B are strikingly similar to
Tycho's remnant.  Both require a hot Fe component and show a radial
segregation of the Fe and Si components of the ejecta. The results of
this analysis indicate that the properties of N103B are consistent
with a Type Ia origin. However, further work must be done,
particularly to determine the location and origin of the O component,
before eliminating a Type II origin.

Certainly, a more realistic 3D model of the ejecta is needed, which
takes into account the initial structure of the remnant, the
expansion, and potentially mixing between the layers. Additionally,
throughout this paper, we have ignored the large asymmetry between the
eastern and western halves of the remnant; to model this remnant more
accurately, this must be taken into account. Using more sophisticated
models for seven radial bins is pointless however, and we suggest that
the analysis be improved by combining this {\it ACIS} dataset with the
0$^{\rm th}$ order {\it Chandra} LETG grating data. With this larger
dataset, the remnant could be sampled with finer radial bins and the
differences between the eastern and western halves could be
explored. Also, by using the information from the {\it Chandra} and
{\it XMM-Newton} gratings observations, one could restrict the
parameters of the O component more effectively, thereby reducing some
of the uncertainties in this analysis.

Finally, this analysis has ignored the intriguing small-scale
variations in brightness and color. An exploration of these may yield
more clues to the origin and distribution of the O and H components.
In particular, it important to determine whether these clumps have the
same composition as the rest of the remnant, or whether they are
dominated by emission lines or continuum.  Again, while several of the
brighter clumps have 10,000 counts, the model we have proposed cannot
be safely used unless a strong Fe K$\alpha$ line is present to remove
some of the confusion between the two different sources of Fe
emission. Again, combining the 0$^{\rm th}$ order {\it Chandra} LETG
grating and {\it ACIS} datasets should improve matters greatly.

\acknowledgements
KTL would like to thank the {\it ACIS} team at Penn State,
particularly, L. Townsley and G. Chartas, for invaluable discussions
about the ACIS instruments and the analysis of {\it Chandra data}. KTL
would also like to thank S. Park and Y. Meada for the many useful
discussions about this project. This research was funded by the {\it
Chandra} grant NAS8-01128. JPH acknowledges partial support from {\it
Chandra} grant GO0-1035X. POS acknowledges partial support from NASA
contract NAS8-39073.

\clearpage

\clearpage

\clearpage

\clearpage

\clearpage

\clearpage

\clearpage

\clearpage

\begin{deluxetable}{lcc}
\tablecaption{\label{ldata}}
\tabletypesize{\footnotesize}
\tablewidth{3.5in}
\tablehead{ 
  \colhead{} & \colhead{He-like Si} & \colhead{S (He- \& H-like)}
          }
\startdata
Line (eV)        & 1750:1915 & 2340:2540 \\
Left Cont. (eV)  & 1565:1730 & 2120:2320 \\
Right Cont. (eV) & 2100:2265 & 2600:2800 \\
Bin Size ($''$)  & 0.5       & 1.0       \\
Smoothing  Scale ($''$)      & 1.5       & 3.0       \\  
\enddata
\tablenotetext{*}{\footnotesize Extraction information for each set
of line images.  Each image was extracted and binned using the {\tt CIAO
2.1} tools. Each image was smoothed to reduce the number of null
pixels without sacrificing the image quality.}
\end{deluxetable}

\clearpage

\begin{deluxetable}{ccc}
\tablewidth{2.4in}
\tablecaption{Radial Extraction Regions \label{ring_table}}
\tabletypesize{\footnotesize}
\tablehead{ 
  \colhead{} & \colhead{Inner} & \colhead{Outer} \\
  \colhead{Ring} & \colhead{Radius ($''$)} & \colhead{Radius ($''$)}
          }
\startdata
1 & ~0.0 & ~5.0 \\
2 & ~5.0 & ~7.0 \\
3 & ~7.0 & ~8.5 \\
4 & ~8.5 & 10.0 \\
5 & 10.0 & 11.0 \\
6 & 11.0 & 12.5 \\
7 & 12.5 & 16.0 \\
\enddata
\end{deluxetable}

\clearpage

\begin{deluxetable}{cl||cl||cl}
\tablewidth{3.0in}
\tablecaption{List of LMC Abundances$^{\dag}$ \label{abund}}
\tabletypesize{\footnotesize}
\tablehead{ }
\startdata
He  &   10.94  $^{*}$  & Mg  &   \,7.08 $^{**}$  & Ca  &   \,5.89 $^{*}$\\
C   &   \,8.04 $^{*}$  & Al  &   \,5.98          & Cr  &   \,5.47 $^{*}$\\
N   &   \,7.14 $^{*}$  & Si  &   \,7.04 $^{**}$  & Fe  &   \,7.01 $^{**}$\\
O   &   \,8.35 $^{*}$  & S   &   \,6.70 $^{*}$   & Co  &   \,4.41        \\
Ne  &   \,7.61 $^{*}$  & Cl  &   \,4.76 $^{*}$   & Ni  &   \,6.04 $^{*}$ \\
Na  &   \,5.83         & Ar  &   \,6.29 $^{*}$   &     &                 \\
\enddata
\tablenotetext{\dag}{\footnotesize Abundances given as 12$\;+\;$log$\;$N(X/H).}
\tablenotetext{*}{\footnotesize Taken from \citet{rd92}}
\tablenotetext{**}{\footnotesize Taken from \citet{H98} since the value from \citet{rd92} 
was more uncertain.}
\end{deluxetable}

\clearpage

\begin{deluxetable}{cllll||ll||ll}
\tablewidth{6.0in}
\tabletypesize{\footnotesize}
\tablecaption{Fit Parameters for the Three Thermodynamic Components $^{*}$ \label{therm_params} }
\tablehead{
   \colhead{    }  & \colhead{    } & \colhead{    } 
                   & \multicolumn{2}{c}{O-Ne-Mg} 
                   & \multicolumn{2}{c}{Si-S-Ar-Ca-Fe} 
                   & \multicolumn{2}{c}{Fe} \\
   \cline{4-9}
   \colhead{    }  & \colhead{     } & \colhead{N$_{H}$}
                   & \colhead{kT}    & \colhead{log n$_{e}$t}
                   & \colhead{kT}    & \colhead{log n$_{e}$t}  
                   & \colhead{kT}    & \colhead{log n$_{e}$t} \\

   \colhead{Ring } & \colhead{$\chi^{2}_{\nu}$/$\nu$} 
                   & \colhead{$10^{22}$cm$^{-2}$}
                   & \colhead{(keV)} & \colhead{(s$\;$cm$^{-3}$)}
                   & \colhead{(keV)} & \colhead{(s$\;$cm$^{-3}$)}
                   & \colhead{(keV)} & \colhead{(s$\;$cm$^{-3}$)} 
          }
\startdata
1 & 1.4/141 & $  0.34^{\; 0.02}_{\; 0.01}$  & $ 1.2^{\;0.2}_{\; 0.1}$  & $11.00^{\; 0.03}_{\; 0.2}$    & $ 1.02^{\; 0.01}_{\;0.01}$   & $\ge 12.9                $ &  $\ge 8$                 & $10.79^{\; 0.02}_{\; 0.09}$\\
2 & 1.2/150 & $  0.36^{\; 0.01}_{\; 0.01}$  & $ 0.87^{\; 0.05}_{\; 0.2}$  & $11.1^{\; 0.2}_{\; 0.2}$  & $ 0.90^{\; 0.02}_{\; 0.02}$  & $\ge 12.6                $ &  $ 4.5^{\; 2}_{\; 0.9}$  & $10.84^{\; 0.03}_{\; 0.07}$ \\
3 & 1.4/147 & $  0.36^{\; 0.01}_{\; 0.01}$  & $ 1.03^{\; 0.06}_{\; 0.06}$ & $11.01^{\; 0.04}_{\; 0.03}$  & $ 1.07^{\; 0.01 }_{\; 0.02}$ & $12.03^{\; 0.2}_{\; 0.07}$ &  $5.2^{\; 0.8}_{\; 0.5}$ & $10.75^{\; 0.02}_{\; 0.02}$\\
4 & 1.3/158 & $  0.40^{\; 0.01}_{\; 0.01}$  & $ 1.07^{\; 0.08}_{\; 0.07}$ & $10.99^{\; 0.04}_{\; 0.2}$ & $ 0.95^{\; 0.06 }_{\; 0.01}$ & $12.4^{\; 0.2}_{\; 0.2}$   &  $6.6^{\; 1}_{\; 0.8}$   & $10.84^{\; 0.01}_{\; 0.1}$  \\
5 & 1.3/155 & $  0.36^{\; 0.01}_{\; 0.02}$  & $ 1.1^{\; 0.3}_{\; 0.1}$    & $11.1^{\; 0.1}_{\; 0.1}$  & $ 0.95^{\; 0.02}_{\; 0.01}$  & $12.3^{\; 0.2}_{\; 0.4}$   &  $ \ge 7 $               & $10.69^{\; 0.02}_{\; 0.02}$ \\
6 & 1.4/159 & $  0.36^{\; 0.02}_{\; 0.01}$  & $ 0.93^{\; 0.2}_{\; 0.04}$ & $11.03^{\; 0.1}_{\; 0.03}$  & $ 0.89^{\; 0.02}_{\; 0.03}$  & $12.1^{\; 0.7}_{\; 0.2}$   &  $ \ge 7$                & $10.70^{\; 0.02}_{\; 0.02}$ \\
7 & 1.7/161 & $ 0.27^{\; 0.02}_{\; 0.02}$  & $ 1.8^{\;0.4}_{\; 0.4}$   & $10.88^{\; 0.07}_{\; 0.1}$   & $ 0.89^{\; 0.02}_{\; 0.02}$  & $\ge 12.4              $   &  $ \ge 2$                & $10.8^{\; 0.2}_{\; 0.1}$    \\
\enddata
\tablenotetext{*}{\footnotesize All errors quoted are 90\%}
\end{deluxetable}

\clearpage

\begin{deluxetable}{cl||lll||lllll||l}
\tablecaption{Fitted Emission Measures $^{*}$ \label{abund_params}}
\tabletypesize{\footnotesize}
\tablewidth{6.5in}
\tablehead{
   \colhead{Ring} & \colhead{H} & \colhead{O} & \multicolumn{2}{c}{EM Ratios} & \colhead{Si} & \multicolumn{4}{c}{EM Ratios} & \colhead{Hot Fe}\\
   \cline{4-5} \cline{7-10}
   \colhead{ } & \colhead{($10^{11}$)$^\dag$} & \colhead{($10^{6}$)$^\dag$} & \colhead{Ne/O} & \colhead{Mg/O} & \colhead{($10^{6}$)$^\dag$} & \colhead{S/Si} & \colhead{Ar/Si} & \colhead{Ca/Si} & \colhead{Fe/Si} &  \colhead{($10^{6}$)$^\dag$ }

          }
\startdata
1 & $ 4.6^{\; 0.2}_{\; 0.4}$    & $ 7.9^{\; 1}_{\; 0.7}$ & $ 1.3^{\; 0.3}_{\; 0.4}$   & $ 0.19^{\; 0.08}_{\; 0.04}$ & $ 6.0^{\; 0.7}_{\; 0.7}$  & $ 0.7^{\; 0.1}_{\; 0.2}$   & $ 0.12^{\; 0.12}_{\; 0.07}$ & $ 0.19^{\; 0.1}_{\; 0.05}$ & $ 0.82^{\; 0.08}_{\; 0.1}$   & $ 2.4^{\; 0.2}_{\; 0.1}$\\
2 & $ 7.0^{\; 0.2}_{\; 0.4}$    & $ 13^{\; 1}_{\; 4}$    & $ 1.0^{\; 0.3}_{\; 0.1}$   & $ 0.24^{\; 0.08}_{\; 0.04}$ & $ 7.0^{\; 0.7}_{\; 0.7}$  & $ 0.9^{\; 0.1}_{\; 0.1}$   & $ 0.28^{\; 0.07}_{\; 0.07}$ & $ 0.4^{\; 0.2}_{\; 0.1}$   & $ \le 0.04$                  & $ 4.5^{\; 0.2}_{\; 0.4}$\\
3 & $ 4.4^{\; 0.1}_{\; 0.1}$    & $ 9.3^{\; 1}_{\; 0.7}$ & $ 1.2^{\; 0.1}_{\; 0.1}$   & $ 0.19^{\; 0.04}_{\; 0.04}$ & $ 7.0^{\;0.3}_{\; 0.3}$   & $ 0.9^{\; 0.1}_{\; 0.1}$   & $ 0.24^{\; 0.05}_{\; 0.1}$  & $ 0.2^{\; 0.1}_{\; 0.1}$   & $ 0.23^{\; 0.05}_{\; 0.05}$  & $ 3.3^{\;0.2}_{\; 0.2}$\\
4 & $ 6.8^{\; 0.2}_{\; 0.2}$    & $ 12^{\; 1}_{\; 2}$    & $ 0.9^{\; 0.1}_{\; 0.1}$   & $ 0.09^{\; 0.03}_{\; 0.02}$ & $ 8.6^{\; 0.7}_{\; 0.7}$  & $ 1.0^{\; 0.1}_{\; 0.1}$   & $ 0.17^{\; 0.1}_{\; 0.07}$  & $ 0.5^{\; 0.1}_{\; 0.1}$   & $ 0.08^{\; 0.1}_{\; 0.02}$   & $ 4.1^{\; 0.1}_{\; 0.2}$\\
5 & $ 4.4^{\; 0.6}_{\; 0.4}$    & $ 9.3^{\; 1}_{\; 0.7}$ & $ 1.1^{\; 0.1}_{\; 0.1}$   & $ 0.17^{\; 0.04}_{\; 0.04}$ & $ 8.0^{\; 0.7}_{\; 0.7}$  & $ 1.0^{\; 0.1}_{\; 0.1}$   & $ 0.26^{\; 0.05}_{\; 0.1}$  & $ 0.4^{\; 0.1}_{\; 0.1}$   & $ 0.23^{\; 0.06}_{\; 0.02}$  & $ 2.8^{\; 0.2}_{\; 0.2}$\\
6 & $ 6.8^{\; 0.6}_{\; 0.4}$    & $ 14^{\; 7}_{\; 2}$    & $ 0.6^{\; 0.1}_{\; 0.4}$   & $ 0.12^{\; 0.02}_{\; 0.02}$ & $ 12.2^{\; 1}_{\; 0.7}$   & $ 1.1^{\; 0.1}_{\; 0.1}$   & $ 0.19^{\; 0.07}_{\; 0.05}$ & $ 0.4^{\; 0.2}_{\; 0.1}$   & $ 0.20^{\; 0.06}_{\; 0.06}$  & $ 2.0^{\; 0.3}_{\; 0.2}$\\
7 & $ 4.4^{\; 0.6}_{\; 0.4}$    & $ 11^{\; 1}_{\; 2}$    & $ 0.6^{\; 0.1}_{\; 0.1}$   & $ 0.11^{\; 0.02}_{\; 0.02}$ & $ 15^{\; 1}_{\; 1}$       & $ 1.0^{\; 0.1}_{\; 0.1}$   & $ 0.34^{\; 0.05}_{\; 0.07}$ & $ 0.3^{\; 0.2}_{\; 0.1}$   & $ 0.23^{\; 0.05}_{\; 0.05}$  & $ 0.6^{\; 0.2}_{\; 0.4}$\\

\enddata

\tablenotetext{*}{\footnotesize All errors quoted are 90\%. The emission 
measures (EMs) are given by $n_{\rm e}n_{\rm i}V/4\pi D^{2}$, where
$n_{\rm e}$ and $n_{\rm i}$ are densities of the electrons and ions,
respectively, $V$ is the volume of the emitting region, and $D$ is the
distance to the SNR. The EMs are in units of cm$^{-5}$}
\tablenotetext{\dag}{\footnotesize The number in parentheses denotes the 
scaling factor. }
\end{deluxetable}

\clearpage

\begin{deluxetable}{cll}
\tablecaption{Model Fits to Emission Measures$^{*}$}\label{3Dfits}
\tabletypesize{\footnotesize}
\tablewidth{2.3in}
\tablehead{ 
   \colhead{} & \colhead{Si} & \colhead{Hot Fe}
          }
\startdata
$\chi ^{2}$         & 20                         & 57                        \\
$\rho_{1} ^{\dag}$  &                            & 0.40$^{\; 0.02}_{\; 0.02}$ \\
$\rho_{2} ^{\dag}$  & 0.51$^{\; 0.03}_{\; 0.05}$ & 0.77$^{\; 0.01}_{\; 0.02}$ \\
K                   & 1.7$^{\; 0.1}_{\; 0.2}$    & 1.12$^{\; 0.07}_{\; 0.07}$ \\
(10$^{7}$cm$^{-5}$) &                            &                            \\
\enddata
\tablenotetext{*}{\footnotesize All errors quoted are 90\%.}
\tablenotetext{\dag}{\footnotesize $\rho$ = R/R$_{tot}$, R$_{tot}$=16 $''$}
\end{deluxetable}

\clearpage

\begin{deluxetable}{ccccc}
\tablecaption{Comparison of Estimated and Model Masses $^{*}$ \label{mass_table}}
\tabletypesize{\footnotesize}
\tablewidth{3.2in}
\tablehead{ 
   \colhead{}        & \colhead{Estimated} & \colhead{Type II $^{\dag}$ }    &\colhead{Type Ia $^{\dag \dag}$}\\
   \colhead{Element} & \colhead{Mass}      & \colhead{13 -- 25 M$_{\odot}$ } &\colhead{W7}
           }
\startdata
Si  & 0.21  & 0.083 & 0.154\\
$\,$S   & 0.22  & 0.029 & 0.085\\
Ar  & 0.07  & 0.005 & 0.015\\
Ca  & 0.11  & 0.004 & 0.012\\
Fe  & 0.34  & 0.099 & 0.626\\
\enddata
\tablenotetext{*}{\footnotesize All masses reported in solar masses.}
\tablenotetext{\dag}{\footnotesize Average of values from \citet{T95}.}
\tablenotetext{\dag \dag}{\footnotesize \citet{I99}.}
\end{deluxetable}

\end{document}